\documentclass{webofc}
\usepackage[varg]{txfonts}   
\usepackage{graphicx}
\usepackage{color}
\def \be  {\begin{equation}}
\def \ee  {\end{equation}}
\def \ee  {\end{equation}}
\def \bea {\begin{eqnarray}}
\def \eea {\end{eqnarray}}
\newcommand{\nn}{\nonumber}
\begin{document}
\title{Quark-hadron phase structure of QCD matter from SU($4$) Polyakov linear sigma model}
%
%

\author{\firstname{Abdel Magied} \lastname{Abdel Aal DIAB}\inst{1,2}\fnsep\thanks{\email{a.diab@eng.mti.edu.eg}} \and
        \firstname{Abdel Nasser} \lastname{TAWFIK}\inst{1,2}\fnsep
        \thanks{\email{a.tawfik@eng.mti.edu.eg}} 
}

\institute{Egyptian Center for Theoretical Physics (ECTP), MTI University, 11571 Cairo, Egypt.
\and
    World Laboratory for Cosmology And Particle Physics (WLCAPP), Cairo, Egypt
          }

\abstract{%
The SU($4$) Polyakov linear sigma model (PLSM) is extended towards characterizing the chiral condensates, $\sigma_l$, $\sigma_s$ and $\sigma_c$  of  light, strange and charm quarks, respectively and the  deconfinement order-parameters $\phi$ and $\bar{\phi}$  at finite temperatures and densities (chemical potentials). The PLSM is considered to study the QCD equation of state  in the presence  of the chiral condensate of charm  for different finite chemical potentials.  The PLSM results are in a good agreement with the recent lattice QCD simulations. We conclude that, the charm condensate is likely not affected by the QCD phase-transition, where  the corresponding critical temperature is greater than that of the light and strange quark condensates.
\label{fig:cndst1}} 
\maketitle

\section{Introduction \label{intro}}

Quantum Chromodynamics (QCD)  is a fundamental theory of the strong interaction acting upon quarks and gluons. Nowadays, the experimental and theoretical studies of charm physics turns to be a very active field in hadronic physics \cite{Brambilla2005}. It was shown that the PLSM model, which is a QCD-like model, describes very successfully the phenomenology of the SU($3$) quark-hadron structure in thermal and dense medium and also at finite magnetic field \cite{OURPLSM201560, OURPLSM201561, OURPLSM201562, OURPLSM201563, OURPLSM201564, OURPLSM201565, OURPLSM201566, OURPLSM201567, OURPLSM201568}. At low densities or as the temperature decreases, the quark gluon plasma (QGP),  where the degrees of freedom of quark flavors, $N_f$, are dominated, is conjectured to confine forming hadrons.  

In the present work, the  PLSM Lagrangian for $N_f$ quark flavours is combined with the Polyakov loops in order to link the chiral symmetry restoration with aspects of the confinement-deconfinement phase transitions. The PLSM Lagrangian is invariant under chiral symmetry $U(N_f)_R\, \times \, U(N_f)_L $ ($N_f=2$ in Ref. \cite{RischkeSU2}, $N_f=3$ in Refs. \cite{OURPLSM201560, OURPLSM201561} and $N_f=4$ in Ref. \cite{SU4ICHEP2016}).  We utilize mean field approximation to the PLSM in order to estimate the pure mesonic potential with $N_f=4$ quark flavours and to describe the dependence of the light, strange and charm quark chiral condensates  $\sigma_l$, $\sigma_s$ and $\sigma_c$ respectively and the Polyakov loop order-parameters $\phi$ and $\bar{\phi}$ in thermal and dense QCD medium. Furthermore, we utilize the PLSM to estimate the well-studied QCD equation-of-state (EoS) and some of other thermodynamic observables. The PLSM calculations are then confirmed to recent lattice QCD simulations.  The paper is organized as follows. In  section (\ref{Model1}), we briefly describe the PLSM.  The results are given in section (\ref{results}), while section (\ref{conclusions}) is devoted to the final conclusions.   

\section{SU($4$) Polyakov linear sigma model  \label{Model1}}

The  PLSM addresses the linkage between the confinement- and deconfinement-phase, where its Lagrangian is composed as $\mathcal{L}_{\mathrm{PLSM}}=\mathcal{L}_{\mathrm{chiral}} -\mathcal{U} (\phi, \phi^*, T)$, and the chiral Lagrangian $\mathcal{L}_{\mathrm{chiral}}= \mathcal{L}_f + \mathcal{L}_m$ is constructed for $N_f$ quark flavours. The Fermionic Lagrangian $\mathcal{L}_f$ reads,
\bea
\mathcal{L}_f = \sum_f \overline{q}_f \left[i\gamma^{\zeta} D_{\zeta}-gT_a(\sigma_a+i \gamma_5 \pi_a)\right] q_f, \label{eq:quarkL}
\eea
and the mesonic Lagrangian, $\mathcal{L}_m$, is given as, 
\bea
\mathcal{L}_m &=&  \mathrm{Tr}(\partial_{\mu}\Phi^{\dag}\partial^{\mu}\Phi-m^2 \Phi^{\dag} \Phi)-\lambda_1 [\mathrm{Tr}(\Phi^{\dag} \Phi)]^2 -\lambda_2 \mathrm{Tr}(\Phi^{\dag}
\Phi)^2 + \mathrm{Tr}[H(\Phi+\Phi^{\dag})] \nn \\ &+& c[\mathrm{Det}(\Phi)+\mathrm{Det}(\Phi^{\dag})]. \label{lmeson}
\eea
The Polyakov-loop potential, $\mathcal{U} (\phi, \bar{\phi}, T)$, introduces color degrees-of-freedom and gluon dynamics with the expectation values of the Polyakov-loop variables $\phi$ and $\bar{\phi}$. There are various functional forms of the Polyakov-loop potential. In the present work, we use the logarithmic form, 
\bea
\frac{\mathcal{U}_{log}(\phi, \bar{\phi}, T)}{T^4} &=& -\frac{1}{2}\, a(T)\, \bar{\phi} \phi + b(T)\, \ln \big[ 1-6  \bar{\phi}  \phi + 4( \bar{\phi}^{3} + \phi^3) - 3 (\bar{\phi}  \phi)^2 \big],
\eea
where the thermal dependent coefficients, $ a(T) = a_0 + a_1 (T_0/T)+a_2 (T_0/T)^2$ and $b(T) = b_3 (T_0/T)^3$, the constants $a_0,\, a_1,\, a_2$ and $b_3$ are fitted to the pure gauge lattice QCD data in \cite{PloyakovLog}.

With basis of light, strange and charm quarks, the  purely mesonic potential can be driven from the mesonic Lagrangian, $\mathcal{L}_m$. The orthogonal basis transformation from the original ones ($\sigma_0, \, \sigma_8$ and $\sigma_{15}$) to the light, the strange and the charm quark flavor basis; i.e. $\sigma_l, \, \sigma_s$ and $\sigma_c$, respectively, can be expressed as
\bea
\sigma_l = \frac{\sigma_0}{\sqrt{2}}  + \frac{\sigma_8}{\sqrt{3}}  + \frac{\sigma_{15}}{\sqrt{6}} , \quad\;\;
\sigma_s = \frac{\sigma_0}{2}- \sqrt{\frac{2}{3}} \sigma_8 + \frac{\sigma_{15}}{2\sqrt{3}} , \quad\;\;
\sigma_c = \frac{\sigma_0}{2}  - \frac{\sqrt{3}}{2} \sigma_{15}. \hspace*{5mm}
\eea
Furthermore, the purely mesonic potential can be obtained as,
\bea
U(\sigma_l, \sigma_s, \sigma_c) &=& - h_l \sigma_l - h_s \sigma_s - h_c \sigma_c  + \frac{m^2\, (\sigma^2_l+\sigma^2_s+\sigma^2_c)}{2} - \frac{c\, \sigma^2_l \sigma_s \sigma_c}{4}  
+ \frac{\lambda_1\, \sigma^2_l \sigma^2_s}{2}  \nonumber \\
&+& \frac{\lambda_1 \sigma_l^2 \sigma_c^2}{2}\, +\, \frac{\lambda_1 \sigma_s^2 \sigma_c^2}{2} + \frac{(2 \lambda_1+\lambda_2)\sigma^4_l }{8}  \, + \,\frac{( \lambda_1 +\lambda_2)\sigma^4_s }{4}+ \frac{(\lambda_1+\lambda_2)\sigma^4_c}{4}.\hspace*{8mm} \label{Upotio}
\eea
In mean-field approximation \cite{Kapusta:2006pm},  all fields are treated as averages in space ($\vec{x}$) and imaginary time ($\tau$). The exchanges between particles and antiparticles shall be expressed by the grand-canonical partition function ($Z$) which can be constructed from  the thermodynamic potential density as 
\bea
\Omega (T, \mu) =-T\, \frac{\ln{Z}}{V}=\, \Omega_{\bar{q}q}(T, \mu) +\,\mathcal{U}_{\mathrm{log}} (\phi, \bar{\phi}, T)+ U(\sigma_l, \sigma_s, \sigma_c). \label{LSMPOT}
\eea 
The potential of the antiquark-quark contribution reads \cite{Fukushima:2008},
\bea 
\Omega_{\bar{q}q}(T, \mu)&=& -2T \sum_{f=l, s, c} \int_0^{\infty} \frac{d^3\vec{P}}{(2 \pi)^3} \left\{ \ln \left[ 1+3\left(\phi+\bar{\phi} e^{-\frac{E_f-\mu}{T}}\right)\times e^{-\frac{E_f-\mu}{T}}+e^{-3 \frac{E_f-\mu}{T}}\right] \right. \nonumber \\ 
&& \hspace*{28mm} \left.  +\ln \left[ 1+3\left(\bar{\phi}+\phi e^{-\frac{E_f+\mu}{T}}\right)\times e^{-\frac{E_f+\mu}{T}}+e^{-3 \frac{E_f+\mu}{T}}\right] \right\}. \hspace*{8mm} \label{PloykovPLSM}
\eea
The subscripts $l$, $s$, and $c$ refer to light, strange and charm quarks, respectively. The energy-momentum dispersion relation is given as $E_f=(\vec{P}^2+m_f^2)^{1/2}$, with $m_f$ being the flavor mass of light, strange, and charm quark coupled to $g$ \cite{blind}; $m_l = g \sigma_l/2$, $m_s=g \sigma_s/\sqrt{2}$ \cite{Kovacs:2006}, and $m_c=g \sigma_c/\sqrt{2}$. In the present work, the Yukawa coupling $g$ is fixed at $g \sim 6.5$. 

In order to evaluate the dependence of the PLSM chiral condensates $\sigma_l$, $\sigma_s$, and  $\sigma_c$ and the deconfinement order-parameters $\phi$ and $\bar{\phi}$ as a function temperature ($T$) and baryon chemical potential $\mu$, the real part of thermodynamic potential, Re($\Omega$) of Eq. (\ref{LSMPOT}), should be minimized at the saddle point
\bea\label{cond1}
\left.\frac{\partial \Omega}{\partial \sigma_l}= \frac{\partial \Omega}{\partial \sigma_s}= \frac{\partial \Omega}{\partial \sigma_c}= \frac{\partial \Omega}{\partial
\phi}= \frac{\partial \Omega}{\partial \bar{\phi}}\right|_{min} =0, \label{PLSMcondition}
\eea

\section{Results  \label{results}}

\begin{figure}[htb] 
\includegraphics[width=4.5cm,angle=-90]{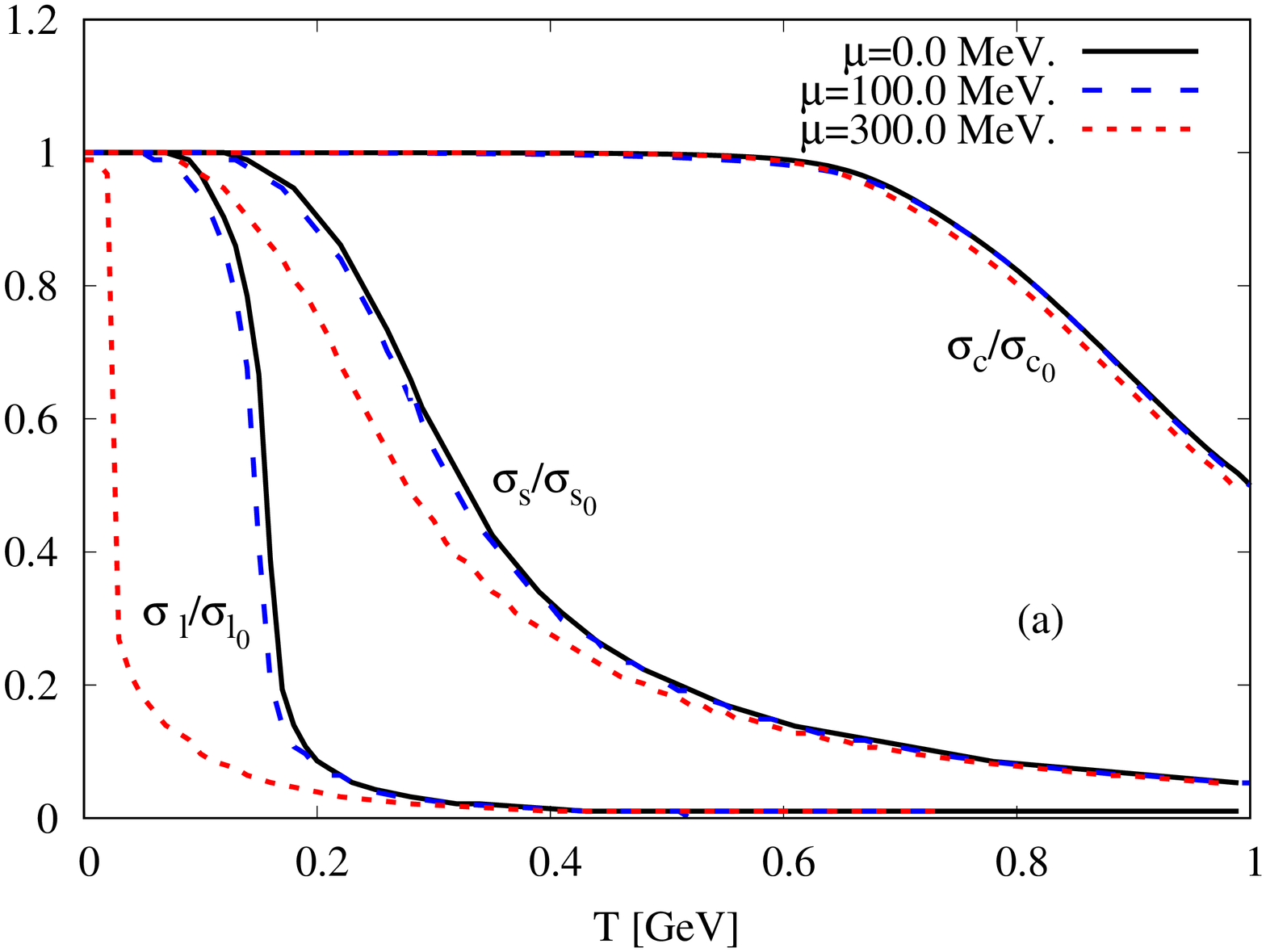}
\includegraphics[width=4.5cm,angle=-90]{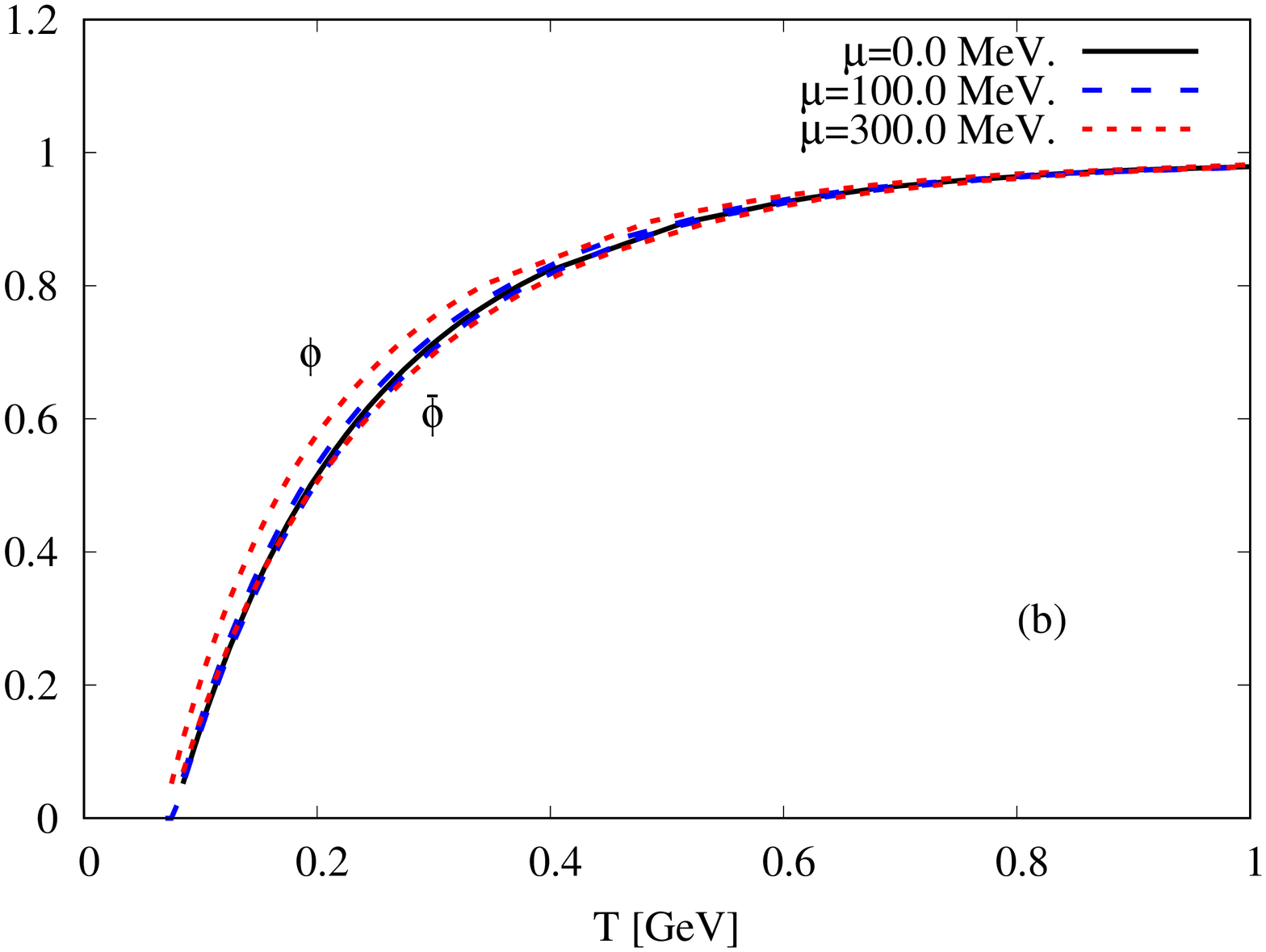}
\caption{In dependence on temperatures, the normalized chiral condensates $\sigma_l$, $\sigma_s$ and $\sigma_c$ normalized with respect to their vacuum values, $\sigma_{l_0}=92.4~$MeV, $\sigma_{s_0}=94.5~$MeV and $\sigma_{c_0}=293.87~$MeV, respectively [left panel (a)] and the order-parameters of Polyakov-loops $\phi$ and $\bar{\phi}$ [right panel (b)] are given at different values of the baryon  chemical potentials $\mu=0,\, 100$ and $300~$MeV  (solid, dashed and dotted curves, respectively). \label{Fig1.ordderPLSM}
} 
\end{figure}

In the following we present the order parameters of the chiral symmetry restoration and the deconfinement phase transitions at finite temperatures and finite baryon chemical potentials. Furthermore, we explore the thermal dependence of some thermodynamic quantities. These results are compared with recent lattice QCD simulations. Fig. \ref{Fig1.ordderPLSM} introduces the thermal dependence of the chiral condensates for light, strange and charm quark flavors, $\sigma _l$, $\sigma _s$ and $\sigma_c$, respectively [left panel (a)], while the order parameters of the deconfinement phase transitions $\phi$ and $\bar{\phi}$ are shown in right panel (b). The PLSM order parameters are given at different values of the baryon chemical potentials $\mu=0$ (solid curves) $100$  (dashed curves) and $300~$MeV (dotted curves), respectively. The quark chiral-condensates are normalized with respect to their vacuum values, $\sigma_{l_0}=92.4~$MeV, $\sigma_{s_0}=94.5~$MeV and $\sigma_{c_0}=293.87~$MeV.

Left panel of Fig. \ref{Fig1.ordderPLSM} illustrates the effect of the baryon chemical potential on the chiral condensates. Whereas the baryon chemical potential increases, the strange and the light condensates move backwards, i.e. the chiral critical temperatures decrease. For the charm condensate, the effect of the baryon chemical potential is absent (not appearing in the temperature range depicted here). Furthermore, the chiral critical temperatures can be obtained from the chiral susceptibility. We find that the charm condensate remains almost temperature independent until the temperature approaches $0.6-0.7~$GeV. It seems that the corresponding critical temperature is much greater than that of the strange and light ones.

Right panel of Fig. \ref{Fig1.ordderPLSM} shows the order-parameters of the Polyakov-loops $\phi$ and $\bar{\phi}$ as functions of temperatures. At $\mu=0~$MeV, we find  that $\phi = \bar{\phi}$, while the Polyakov loop variables are different at finite baryon chemical potentials. We find that with increasing the baryon chemical potentials, the order-parameters of Polyakov-loops $\phi$ moves backwards but $\bar{\phi}$ moves towards, i.e. the deconfinement critical temperatures of $\phi$ decrease but that of $\bar{\phi}$ increase with the increase in  baryon  chemical potentials.

It is believed that the thermodynamic quantities are sensitive to the change accompanying the confinement-deconfinement phase transitions. Some of the thermodynamic quantities such as pressure, entropy density, energy and interaction measure, are given as $p=-\Omega(T,\,\mu),\; s=\partial p/\partial T,\; \epsilon=-p + Ts$ and $(\epsilon-3p)/T^4$, respectively. In the present work, we focus on comparing our PLSM results on the thermal dependence of normalized pressure and of interaction measure with recent lattice QCD simulation \cite{Lattice_QCD}.

\begin{figure}[htb] 
\includegraphics[width=4.5cm,angle=-90]{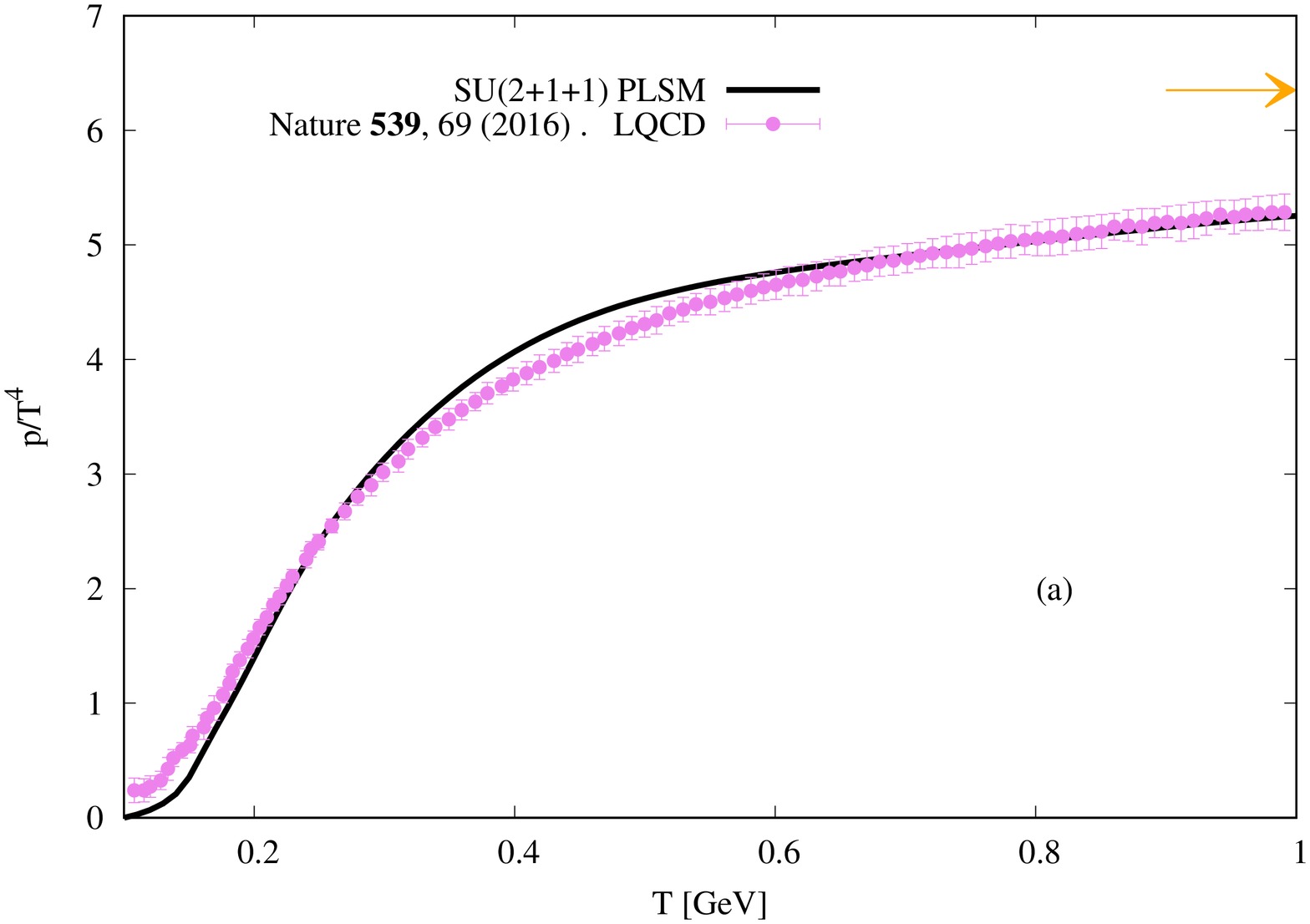}
\includegraphics[width=4.5cm,angle=-90]{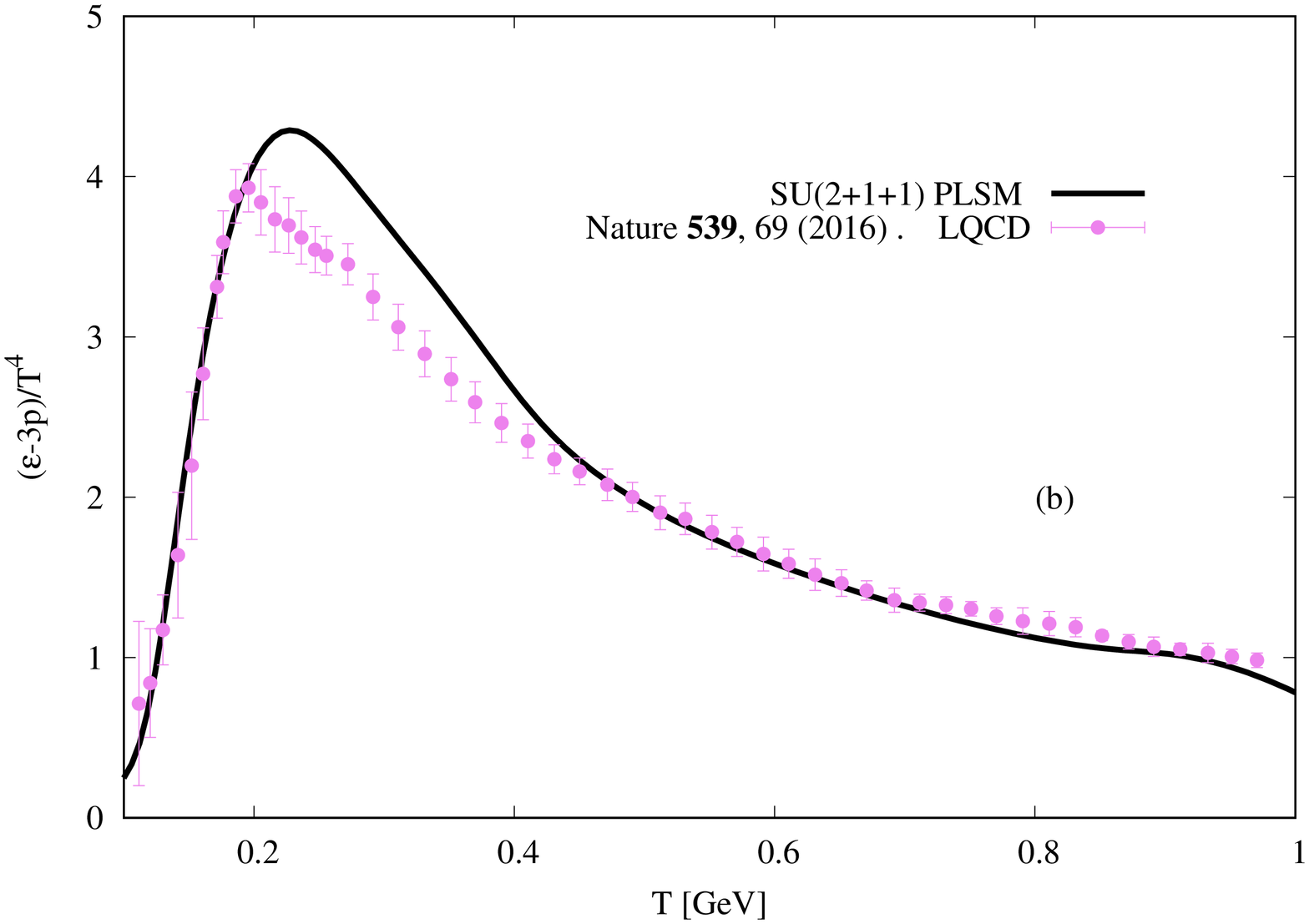}
\caption{ The normalized pressure [left panel (a)] and the the interaction measure [right panel (b)] are depicted in dependence on temperature. The PLSM calculations (solid curves) are compared with recent lattice QCD simulations (symbols)  \cite{Lattice_QCD}.
\label{fig2:ThermoPLSM}} 
\end{figure}

Fig. \ref{fig2:ThermoPLSM} shows the thermal dependence of the normalized pressure [left panel (a)] and the interaction measure [right panel (b)] at vanishing baryon chemical potential. The PLSM calculations are confronted to recent lattice QCD simulations  \cite{Lattice_QCD}.  It is obvious that the normalized pressure increases with increasing temperatures and saturates at high temperatures. We observe that the pressure doesn't exceed the Stefan-Boltzmann (SB) limit at $N_f=4$, i.e. massless quarks and gluons, 
\bea
\frac{P_{SB}}{T^4} = \frac{\pi^2}{45} (N_c^2 -1 )+ N_c N_f \left[ \frac{7 \pi^2}{180} +\frac{1}{6} \left(\frac{\mu}{T}\right)^2 + \mathcal{O}\left(\frac{\mu}{T}\right)^4\right], 
\eea
where $N_c$ are the color dof, the first term refers to the contributions of gluons and the second one to the  quarks. The value of thermodynamic pressure is low in confined phase and begins to rise in the deconfined phase, when temperature increases. In all calculations such as the lattice QCD simulations \cite{Lattice_QCD} and even other non-perturbative approaches such as \cite{Schaefer2010}, the normalized pressure stays below the SB limit. In out calculations as well. At vanishing chemical potential, the pressure in dependence on temperature behaves smoothly through the smooth crossover. At temperatures as much as two times the critical value, the pressure reaches approximately $80\%$ of the SB limit. In both hadronic (low temperatures) and tQGP (high temperatures) regions, a good agreement with recent lattice QCD calculations \cite{Lattice_QCD} is obtained. 

The interaction measure or the trace anomaly of the energy-momentum tensor  $\mathcal{T^{\mu \nu}}$,
\bea
\Delta= \frac{\mathcal{T^{\mu\mu}}}{T^4} = \frac{\epsilon-3p}{T^4 }= T \frac{\partial}{\partial T}(p/T^4),
\eea
is a  dimensionless thermodynamic quantity depicted in right panel of Fig. \ref{fig2:ThermoPLSM} at vanishing baryon chemical potential and compared with recent lattice QCD calculations.  We observe that the interaction measure changes smoothly around $T_c$, where the interacting system is conjectured to experience crossover. The thermal dependence of the interaction measure appears to agree well with the recent lattice QCD calculations \cite{Lattice_QCD}, in both hadronic (low temperatures) and QGP (high temperatures) phases. In the broken phase, it shows a peak around $T_c$, but a rapid decrease is seen in the restored phase. 

\section{Conclusion \label{conclusions}}

A systematic study for the chiral phase-structure of heavy mesons and their decay widths in thermal and dense medium is very closely connected with recent theoretical, computational and experimental results. Thus, we present an extension to the PLSM to $N_f=4$ quarks flavors. Accordingly, the extra degrees-of-freedom modifies the PLSM thermodynamic potential $\Omega(T,\mu)$. We present the thermal dependence of the light, strange and charm quark chiral condensates, $\sigma_l$, $\sigma_s$ and $\sigma_c$, respectively and the deconfinement order-parameters ($\phi$ and $\bar{\phi}$) at different baryon chemical potentials. For the charm condensate, we notice that its sensitivity to the baryon chemical potentials and temperatures is very weak until  $0.6- 0.7~$GeV. We also observe a fair agreement with the recent lattice QCD calculations for the thermodynamic pressure and the interaction measure.


\begin{thebibliography}{}
%
%
\bibitem{Brambilla2005} N. Brambilla, A. Pineda, J. Soto and A. Vairo, Rev. Mod. Phys. {\bf 77}, 1423 (2005).

\bibitem{OURPLSM201560} A. Tawfik, N. Magdy, and A. Diab, Phys. Rev. C {\bf 89}, 055210 (2014).

\bibitem{OURPLSM201561} Abdel Nasser Tawfik, Abdel Magied Diab, Phys. Rev. C {\bf 91}, 015204 (2015).
 
\bibitem{OURPLSM201562} Abdel Nasser Tawfik, Abdel Magied Diab, and M.T. Hussein, 
{\it ''Quark-hadron phase structure and QCD equations of state in vanishing and finite magnetic field''}, 
1604.08174 [hep-lat].
  
 \bibitem{OURPLSM201563} Abdel Nasser Tawfik, Abdel Magied Diab, and M. T. Hussein, Int. J. Adv. Res. Phys. Sci. {\bf 3}, 4 (2016).
   
\bibitem{OURPLSM201564} Abdel Nasser Tawfik, Abdel Magied Diab, and M. T. Hussein,  Int. J. Mod. Phys. A {\bf 31}, 1650175 (2016).
     
\bibitem{OURPLSM201565} Abdel Nasser Tawfik, Abdel Magied Diab, N. Ezzelarab, and A. G. Shalaby, 
Adv. High Energy Phys. {\bf 2016},  1381479 (2016).

\bibitem{OURPLSM201566} A. Tawfik and N. Magdy, J. Phys. G {\bf 42}, 015004 (2015).

\bibitem{OURPLSM201567} A. Tawfik and N. Magdy, Phys. Rev. C {\bf 91}, 015206 (2015).
 
\bibitem{OURPLSM201568} A. Tawfik and N. Magdy, Phys. Rev. C {\bf 90}, 015204 (2014).

\bibitem{RischkeSU2} D. Parganlija, F. Giacosa and D. H. Rischke, Phys. Rev. D {\bf 82}, 054024 (2010).

\bibitem{SU4ICHEP2016} Abdel Magied Diab, Azar I. Ahmadov, Abdel Nasser Tawfik, Eiman Abou El Dahab, PoS {\bf ICHEP2016}, 634 (2016).

\bibitem{PloyakovLog} S. Roessner, C. Ratti, and W. Weise, Phys. Rev. D {\bf 75}, 034007 (2007).

\bibitem{Kapusta:2006pm} J. I. Kapusta and C. Gale, {\it ''Finite-temperature field theory: Principles and applications''},  (Cambridge University Press, UK, 2006).

\bibitem{Fukushima:2008} K. Fukushima, Phys. Rev. D {\bf 77}, 114028 (2008).

\bibitem{blind} O. Scavenius, A. Mocsy, I. N. Mishustin, and D. H. Rischke, Phys. Rev. C {\bf 64}, 045202 (2001). 

\bibitem{Kovacs:2006} P. Kovacs and Z. Szep, Phys. Rev. D {\bf 75}, 025015 (2007).  
  
\bibitem{Lattice_QCD}  S. Borsanyi, {\it et. al.} "{\it Lattice QCD for cosmology}", Nature {\bf 539}, 69 (2016). 
  
\bibitem{Schaefer2010}  B-J Schaefer, M. Wagner, J. Wambach Phys. Rev. D {\bf 81}, 074013 (2010).


	
 

%
\end{thebibliography}
\end{document}